\documentclass[prd,superscriptaddress,showpacs,twocolumn,footinbib]{revtex4}

\usepackage{amsmath}
\usepackage{slashed}
\DeclareMathOperator{\tr}{tr}
\DeclareMathOperator{\diag}{diag}

\begin{document}

\title{Color superconductor with a color-sextet condensate}
\author{Tom\'a\v s Brauner}
\email{brauner@ujf.cas.cz}
\affiliation{Dept. Theoretical Physics, Nuclear Physics Institute, 25068 \v Re\v z (Prague), Czech Republic}
\affiliation{Faculty of Mathematics and Physics, Charles University, Prague, Czech Republic}
\author{Ji\v r\'{\i} Ho\v sek}
\affiliation{Dept. Theoretical Physics, Nuclear Physics Institute, 25068 \v Re\v z (Prague), Czech Republic}
\author{Rudolf S\'ykora}
\affiliation{Faculty of Mathematics and Physics, Charles University, Prague, Czech Republic}

\begin{abstract}
We analyze color superconductivity of one massive flavor quark matter at
moderate baryon density with a spin-zero color-sextet condensate. The most
general Higgs-type ground-state expectation value of the order parameter
implies complete breakdown of the $SU(3)\times U(1)$ symmetry. However, both
the conventional fourth-order polynomial effective bosonic description, and the
NJL-type fermionic description in the mean-field approximation favor an
enhanced $SO(3)$ symmetry of the ground state. We ascribe this finding to the
failure of the mean-field approximation and propose that a more sophisticated
technique is needed.
\end{abstract}

\maketitle

\section{Introduction}
Viewing the low-temperature deconfined QCD matter at moderate baryon densities
as a BCS-type color superconductor is based on good assumptions (see
\cite{Barrois:1977xd,Frautschi:1978rz,Bailin:1984bm} for original references
and \cite{Rajagopal:2000wf} for a recent review). First, the only degrees of
freedom relevant for the effective field theory description of such a matter
are the relativistic colored quark fields with their appropriate Fermi
surfaces. The colored gauge fields can be introduced perturbatively, and
eventually switched off in the lowest approximation. Second, the quarks
interact with each other by an attractive interaction providing for Cooper
instability. It is natural to speak of the Higgs phases of QCD
\cite{Alford:2002ng}.

Due to the mere fact that the quarks carry the Lorentz index (spin), color and
flavor, the ordered colored-quark phases could be numerous. Which of them is
energetically most favorable depends solely upon the numerical values of the
input parameters (chemical potentials, and the dimensionful couplings) in the
underlying effective Lagrangian. Because there are no experimental data on the
behavior of the cold deconfined quark matter available, all generically
different, theoretically safe \cite{Sannino:2000kg} and interesting
possibilities should be phenomenologically analyzed. Moreover, one should be
prepared to accept the fact that one or both our assumptions can be invalid. In
any case there are the low-temperature many-fermion systems which are not the
Landau--Fermi liquids, and which become peculiar superconductors
\cite{Polchinski:1992ed}.

Recently, all distinct forms of the quasiquark dispersion laws corresponding
to different sets of 16 matrices in the Lorentz index were
systematically derived \cite{Alford:2002rz}. Those exhibiting
spontaneous breakdown of the
rotational symmetry manifested in the anisotropic form of the dispersion
law are particularly interesting. Their possible nodes can yield
important physical consequences even if the corresponding gaps are
numerically small \cite{Buballa:2002wy}.

To have a complete list of different ordered quantum phases of the quark matter
it would be good to know what is the pattern of spontaneous breakdown of the
color $SU(3)$ if an effective interaction prefers not the standard quark-quark
Cooper pairing in the antisymmetric color antitriplet, but rather in the
symmetric color sextet. Such a pairing would influence qualitatively not only
the quark, but also the gluon spectrum.

Although the explicit analysis presented in this paper is strictly
phenomenological we describe here briefly a mechanism which, within QCD and
under plausible assumptions, can yield the desired color-sextet diquark
condensate. Instabilities of the perturbative QCD in the two-gluon channel
discussed in \cite{Hansson:1982dv} justify contemplating several types of
effective colored excitations in the deconfined phase at moderate densities
with effective (but in practice theoretically unknown) couplings to both quarks
and gluons. According to \cite{Hansson:1982dv}, there should be four types of
two-gluon collective excitations: spin-zero color-singlet, spin-zero color
octet, spin-one color octet, and spin-two color 27-plet. It is easy to show
that exchange of a massive color-octet scalar results in a four-quark
interaction
\begin{equation}
\label{ScalarExchangeInt}
{\cal L_{\text{int}}}=G(\bar\psi\vec\lambda\psi)^2,
\end{equation}
with $G>0$, which is necessary for the color-sextet diquark condensation. It
is, however, not easy to show which of the exchanges, including the one-gluon
one, is eventually the most important. In fact, exchange of the color-singlet
scalar would also lead to an attractive interaction in the color-sextet
quark--quark channel, but as we aim at a phenomenological analysis and do not
attempt to evaluate the effective coupling $G$, we restrict ourselves in the
following to the single interaction term \eqref{ScalarExchangeInt}.

We note that the argument leading to the conjectured colored collective modes
excited by two gluon operators is the same as that leading, in the quark
sector, to the phenomenologically useful \cite{Jaffe:2003sg} color-antitriplet
scalar field with the quantum numbers of a diquark.

The possibility of diquark condensation in the color-symmetric channel has
already been investigated in various contexts, for instance, within the
color-flavor-locking scheme \cite{Alford:1998mk}, and as an admixture to the
color-antitriplet condensate \cite{Schafer:1999fe,Giannakis:2001wz}. The
algebraic structure of spontaneous symmetry breaking due to an $SU(3)$-sextet
condensate is, however, richer than so far discussed in literature, and it is
the general characterization of this structure that we focus on here.

The outline of the paper is as follows. In Sec. II we describe the color-sextet
superconductivity phenomenologically i.e., in terms of a scalar color-sextet
Higgs field. We are not aware of any systematic treatment of the Higgs
mechanism with an $SU(3)$ sextet in the literature and, therefore we go quite
into detail. In Sec. III we review main ideas of the semi-microscopic approach
i.e., a self-consistent BCS-type approximation for a relativistic fermionic
second-quantized quark field, and apply it to the case of color-sextet
condensation. Section IV contains a summary and a brief discussion of the
obtained results, and comparison of the two approaches.

\section{Higgs mechanism with an $SU(3)$ sextet}
Simplifying as much as possible we consider the relativistic quark matter of
one massive flavor (say $s$-quark matter) in the deconfined phase at moderate
baryon density. We assume that its ground state is characterized by the
quark-quark Cooper-pair condensate in the antisymmetric spin zero state. By
Pauli principle this means the symmetric sextet state in $SU(3)$ i.e.,
\begin{equation}
\label{condensate}
 \langle0|\psi_{\alpha i}(C\gamma_5)_{\alpha \beta}
 \psi_{\beta j}|0\rangle\propto\langle\Phi_{ij}\rangle_{0},
\end{equation}
where we insert a dimensionful constant of proportionality to make $\Phi$ a
dimension-one operator. The constant of proportionality can be determined
within the mean-field approximation to be $3/2G$, see Sec. III.

Treating the $u$ and $d$ quarks as nearly degenerate in mass and both much
lighter than the $s$ quark, such a condensate may provide a complement to the
usual picture of $u$ and $d$ pairing in the color-antitriplet channel
\cite{Alford:1999pa}.

In an effective Higgs description $\Phi_{ij}$ is a spin-zero color-sextet order
parameter which transforms under the color $SU(3)$ as a complex symmetric
matrix,
\[
\Phi\to U\Phi U^T.
\]
The dynamics of $\Phi$ is governed by the most general Lagrangian invariant
under global $SU(3)\times U(1)$ and spacetime transformations. As the full
Lorentz invariance is explicitly broken by the presence of a dense medium, we
require that the Lagrangian be invariant under spatial rotations only.

Since we aim at an effective description of the superconducting phase,
renormalizability is not an issue here, and we have to include all possible
interactions built up from the sextet $\Phi$ that respect the symmetry of the
theory.

In accordance with our assumptions, the gauge interaction can be switched on
perturbatively by gauging the global $SU(3)$ color symmetry. Formally, we just
replace the ordinary derivative of $\Phi$ with the covariant derivative
\begin{equation}
D_{\mu}\Phi = \partial_{\mu} \Phi -ig A_{\mu}^{a}
\left(\frac{1}{2}\lambda_{a} \Phi + \Phi \frac{1}{2}\lambda^{T}_{a}\right),
\label{cov_der}
\end{equation}
where $A^a_{\mu}$ is the colored gluon field. The effective Lagrangian thus has
the form
\begin{equation}
{\cal L} = \alpha_e\tr(D_0\Phi)^{\dagger}D^0\Phi + \alpha_m\tr(D_i\Phi)^{\dagger}D^i\Phi
           - V(\Phi) + \dots,
\label{L_GL}
\end{equation}
where $V(\Phi)$ is the most general $SU(3)\times U(1)$-invariant polynomial in
$\Phi$ and the ellipses stand for other possible terms that involve
covariant derivatives and/or gauge field strength tensors $F_{a\mu\nu}$.

\subsection{$SU(3)$ invariants from a sextet}
The ground-state expectation value $\langle\Phi\rangle_{0}=\phi$ is at the tree
level given by the minimum of the scalar potential $V(\Phi)$. To proceed with
our analysis, we have to specify its concrete form.

Note that the group $SU(3)$ has only three algebraically independent invariant
tensors, namely $\delta^i_j$, $\varepsilon_{ijk}$, and $\varepsilon^{ijk}$, the
lower and upper indices transforming under the fundamental representation of
$SU(3)$ and its complex conjugate, respectively (see, for example,
\cite{Georgi:1982jb}). As a consequence, the most general $SU(3)\times U(1)$
invariant built up from a single sextet $\Phi$ can be constructed from products
and sums of $\det(\Phi^{\dagger}\Phi)$ and $\tr(\Phi^{\dagger}\Phi)^n$, the
symbols ``$\det$'' and ``$\tr$'' referring to determinant and trace in the
color space, respectively \footnote{ The $U(1)$ symmetry guarantees that the
$SU(3)$ invariants $\det\Phi$ and $\det\Phi^{\dagger}$ always come with the
same power.}.

Of these polynomials, however, only three are algebraically independent.
Indeed, express
\begin{align*}
\tr \Phi^{\dagger} \Phi &= \alpha + \beta + \gamma, \\
\tr (\Phi^{\dagger} \Phi)^2 &= \alpha^2 + \beta^2 + \gamma^2, \\
\det \Phi^{\dagger} \Phi &= \alpha\beta\gamma,
\end{align*}
where $\alpha,\beta,\gamma$ are the eigenvalues of $\Phi^{\dagger}\Phi$
\footnote{Here and in the following, we act as if $\Phi^{\dagger}\Phi$ were a
c-number matrix and not an operator one. We can do so since all components of
$\Phi^{\dagger}\Phi$ commute with one another. The constants
$\alpha,\beta,\gamma$ are to be understood as eigenvalues of the $3\times3$
matrix $\Phi^{\dagger}\Phi$, disregarding its operator nature. Our goal is the
generating function \eqref{GeneratingFunction} which yields purely algebraic
relations among $\det(\Phi^{\dagger}\Phi)$ and $\tr(\Phi^{\dagger}\Phi)^n$,
without any reference to $\alpha,\beta,\gamma$. Our conclusions about
(in)dependence of various polynomials are thus valid for c-number as well as
operator matrices.}, and define the symmetric polynomials
\begin{align*}
\pi_1 &= \alpha + \beta + \gamma, \\
\pi_2 &= \alpha\beta+\alpha\gamma+\beta\gamma =
\frac{1}{2} [-\tr (\Phi^{\dagger} \Phi)^2 + (\tr \Phi^{\dagger} \Phi)^2], \\
\pi_3 &= \alpha\beta\gamma.
\end{align*}
Note that the values of $\pi_1,\pi_2,\pi_3$ determine those of
$\alpha,\beta,\gamma$ uniquely as the three roots of the cubic equation
$x^3-\pi_1x^2+\pi_2x-\pi_3=0$. Thus also the values of all $\tr (\Phi^{\dagger}
\Phi)^{n} = \alpha^{n} + \beta^{n} + \gamma^{n}$ for $n \geq 3$ are fixed.
Moreover, they can be expressed directly in terms of $\pi_1,\pi_2,\pi_3$ as the
Taylor coefficients of the generating function
\[
f(t)\equiv\tr\ln(1+t\Phi^{\dagger}\Phi)=\ln\det(1+t\Phi^{\dagger}\Phi),
\]
which is readily rewritten as
\begin{equation}
\label{GeneratingFunction}
f(t)=\ln(1+\pi_1t+\pi_2t^2+\pi_3t^3).
\end{equation}

We have thus shown that the scalar potential $V(\Phi)$ can always be expressed
as a function of the three independent invariants $\det(\Phi^{\dagger}\Phi)$,
$\tr(\Phi^{\dagger}\Phi)$, and $\tr(\Phi^{\dagger}\Phi)^2$.

\subsection{Symmetry-breaking patterns}
We shall now turn to the structure of the ground state. In our effective Higgs
approach, the $SU(3)\times U(1)$ symmetry is spontaneously broken by the
ground-state expectation value $\phi$ of the field $\Phi$, which is a constant
due to the translation invariance of the ground state. We can exploit the
symmetry to give the $\phi$ as simple form as possible. In fact, as shown by
Schur \cite{Schur:1945ab}, any complex symmetric matrix can always be written
as
\[
\phi=U\Delta U^{T},
\]
where $U$ is an appropriate unitary matrix, and $\Delta$ is a real, diagonal
matrix with non-negative entries. In our case, we set $\Delta=\diag(\Delta_{1},
\Delta_{2}, \Delta_{3})$.

Consequently, there are several distinct patterns of spontaneous symmetry
breaking possible.

(a) $\Delta_{1}>\Delta_{2}>\Delta_{3}>0$. This ordering can always be achieved
by the allowed appropriate real orthogonal transformations. The continuous
$SU(3)\times U(1)$ symmetry is completely broken (only a discrete $(Z_2)^3$
symmetry is left).

(b) Two $\Delta$'s are equal, say $\Delta_{1}=\Delta_{2}\neq\Delta_{3}$. This
implies an enhanced $O(2)$ symmetry in the corresponding $2\times2$ block of
$\phi$.

(c) $\Delta_{1}=\Delta_{2}=\Delta_{3}\neq0$. The vacuum remains $O(3)$ symmetric.

(d) Some of $\Delta_{i} = 0$. Then there is a residual $U(1)$ or $U(2)$
symmetry of the vacuum corresponding to the vanishing entry or entries of
$\Delta$.

The concrete type of the symmetry breaking pattern is determined by the scalar
potential $V(\Phi)$. Note that, having relaxed the renormalizability
requirement, we can always choose the potential $V(\Phi)$ so that it yields as
its minimum any desired values of $\Delta_1,\Delta_2,\Delta_3$, just take
\begin{multline*}
V(\Phi) = \frac 12 a_1 \left[\tr\Phi^{\dagger}\Phi - \pi_1 \right]^2 + \\
+ \frac 12 a_2 \left[\tr(\Phi^{\dagger}\Phi)^2 - \pi_{1}^{2} + 2\pi_{2}
\right]^{2} + \frac 12 a_3 \left[\det\Phi^{\dagger}\Phi - \pi_{3} \right]^{2}
\label{V}
\end{multline*}
with all $a_1,a_2,a_3$ positive. The $\pi$'s here are to be interpreted as
vacuum expectation values of the corresponding operators.

\subsection{Higgs mechanism with a quartic potential}
Up to now we have repeatedly stressed the fact that we are dealing with an
effective theory and therefore we should include in our Lagrangian all possible
interactions preserving the $SU(3)\times U(1)$ symmetry.

Nevertheless, under some specific conditions it is plausible to start up with a
renormalizable linear sigma model that is, take a general quartic potential
$V(\Phi)$ and neglect all operators of dimension greater than four. In Sec. IV
we will see that this rather restrictive choice is justified when the
underlying microscopic interaction is of four-fermion type.

We thus take up a general quartic potential \footnote{Similar analyses within a
more general class of models have been performed in
\cite{Paterson:1981fc,Pisarski:1981ps}.},
\begin{equation}
\label{QuarticPotential}
V(\Phi)=-a\tr\Phi^{\dagger}\Phi+b\tr(\Phi^{\dagger}\Phi)^2+c(\tr\Phi^{\dagger}\Phi)^2,
\end{equation}
where the minus sign at $a$ suggests spontaneous symmetry breaking at the tree
level. Varying \eqref{QuarticPotential} with respect to $\Phi^{\dagger}$, we
derive a necessary condition for the vacuum expectation value $\phi$,
\begin{equation}
\label{MinimumOfV}
-a\phi+2b\phi\phi^{\dagger}\phi+2c\phi\tr(\phi^{\dagger}\phi)=0.
\end{equation}
A simple observation of \eqref{MinimumOfV} reveals that, should the matrix
$\phi$ be non-singular, we can divide by it and arrive at the condition
\[
2b\phi^{\dagger}\phi=a-2c\tr(\phi^{\dagger}\phi).
\]
Thus, unless $b=0$, $\phi^{\dagger}\phi$ and hence also $\Delta$ must be
proportional to the identity matrix.

Moreover, even when $\phi$ is singular, it can be replaced with the real
diagonal matrix $\Delta$ and we see from \eqref{MinimumOfV} that all non-zero
entries $\Delta_i$ satisfy the equation
\[
2b\Delta_i^2=a-2c\tr\Delta^2.
\]
Thus all non-zero $\Delta$'s develop the same value.

Which of the suggested solutions of \eqref{MinimumOfV} represents the absolute
minimum of the potential depends on the input parameters $a,b,c$, which must be
inferred from the underlying theory \footnote{Without further knowledge, we can
only constrain the values of $b$ and $c$ by the requirement of boundedness of
$V(\Phi)$ from below. It is clear that at least one of these parameters must be
positive, positivity of both being, of course, the safest choice. The sizes of
the two quartic interaction terms are restricted by the inequalities
$\tr(\Phi^{\dagger}\Phi)^2\leq(\tr\Phi^{\dagger}\Phi)^2\leq3\tr(\Phi^{\dagger}\Phi)^2$
where the equality sign in the left and right hand side inequality occurs when
only one $\Delta_i$ is non-zero and $\Delta_1=\Delta_2=\Delta_3$, respectively.
The potential is thus bounded from below if and only if, $b$ is positive and
$c>-b/3$, or $c$ is positive and $b>-c$.}. We therefore stop the Higgs-like
analysis here with the simple conclusion that under fairly general
circumstances the quartic potential can be minimized by a matrix $\Delta$
proportional to the unit matrix, thus leading to an interesting
symmetry-breaking pattern (see the paragraphs (c) above and below).

\subsection{Gluon mass spectrum}
Let us now switch on the gauge interaction perturbatively. Due to the
spontaneous symmetry breaking some of the gluons acquire non-zero masses via
the Higgs mechanism. At the lowest order of the power expansion in the
effective theory, the mass matrix of the gluons follows from the scalar field
kinetic terms in \eqref{L_GL} upon replacing $\Phi$ with $\phi$.

Now, recalling the particular form of the covariant derivative in
\eqref{cov_der}, we arrive at the following gluon mass squared matrix:
\begin{widetext}
\begin{multline*}
M_{e,m}^2 = \alpha_{e,m}g^2\times \\
\left[
\begin{array}{cccccccc}
(\Delta_1+\Delta_2)^2 & 0 & 0 & 0 & 0 & 0 & 0 & 0\\
0 & (\Delta_1-\Delta_2)^2 & 0 & 0 & 0 & 0 & 0 & 0\\
0 & 0 & 2(\Delta_1^2+\Delta_2^2) & 0 & 0 & 0 & 0 & \frac{2}{\sqrt3}(\Delta_1^2-\Delta_2^2)\\
0 & 0 & 0 & (\Delta_1+\Delta_3)^2 & 0 & 0 & 0 & 0\\
0 & 0 & 0 & 0 & (\Delta_1-\Delta_3)^2 & 0 & 0 & 0\\
0 & 0 & 0 & 0 & 0 & (\Delta_2+\Delta_3)^2 & 0 & 0\\
0 & 0 & 0 & 0 & 0 & 0 & (\Delta_2-\Delta_3)^2 & 0\\
0 & 0 & \frac{2}{\sqrt3}(\Delta_1^2-\Delta_2^2) & 0 & 0 & 0 & 0 &
\frac23(\Delta_1^2+\Delta_2^2+4\Delta_3^2)
\end{array}
\right]
\end{multline*}
\end{widetext}
The subscripts $e,m$ distinguish between the temporal (``electric'') and
spatial (``magnetic'') components of the gluon field.

Let us briefly comment on the above mentioned four types of symmetry breaking
patterns.

(a) $\Delta_{1}>\Delta_{2}>\Delta_{3}>0$. The $SU(3)\times U(1)$ symmetry is
completely broken, therefore there are nine massless Nambu--Goldstone modes.
Eight of them are eaten by the gluons, which thus acquire non-zero unequal
masses (with an appropriate diagonalization in the $(A^3,A^8)$ block). There is
one physical Nambu--Goldstone boson corresponding to the broken global $U(1)$
baryon number symmetry of the underlying theory. Going to the unitary gauge, we
can transform away eight of the original twelve degrees of freedom and
parameterize the sextet field $\Phi$ as
\[
\Phi(x)=\frac{1}{\sqrt2}e^{i\theta(x)}\left(
\begin{array}{ccc}
\Delta_1(x) & 0 & 0\\
0 & \Delta_2(x) & 0\\
0 & 0 & \Delta_3(x)
\end{array}
\right),
\]
the $\Delta$'s representing three massive radial modes and $\theta$ the
Nambu--Goldstone mode.

(b) $\Delta_1=\Delta_2\neq\Delta_3$. One gluon is left massless, corresponding
to the Gell-Mann matrix $\lambda_2$ which generates the $SO(2)$ symmetry of the
ground state.

(c) $\Delta_1=\Delta_2=\Delta_3\neq0$. There are three massless gluons
corresponding to the generators $\lambda_2,\lambda_5,\lambda_7$ of the $SO(3)$
subgroup of $SU(3)$. All other gluons receive equal masses so that the symmetry
breaking $SU(3)\to SO(3)$ is isotropic.

(d) Some of $\Delta_{i} = 0$. There is always an unbroken global $U(1)$
symmetry that arises from a combination of the original baryon number $U(1)$
and the diagonal generators of $SU(3)$, hence all Nambu--Goldstone modes that
stem from the symmetry breaking are absorbed into the gauge bosons.

\subsection{Interpretation of the results}
So far in this section, we have worked out the usual Higgs mechanism for the
case that the scalar field driving the spontaneous symmetry breaking transforms
as a sextet under the color $SU(3)$. However, one must exercise some care when
applying the results to the physical situation under consideration, that is,
color superconductivity. In the very origin of possible problems lies the fact
that $\Phi$ is not an elementary dynamical field but rather a composite order
parameter.

Anyway, our analysis of symmetry breaking patterns still holds as for this
purpose one can regard $\Phi$ as simply a shorthand notation for the condensate
in Eq. \eqref{condensate}.

The most apparent deviation from the standard Higgs mechanism is the presence
of non-trivial normalization constants at the kinetic terms in \eqref{L_GL}.
This is due to the compositeness of the field $\Phi$
\cite{Pisarski:1999gq,Rischke:2000qz}.

Further, the power expansion of the effective Lagrangian \eqref{L_GL} can be
reliable as long as the expansion parameter is sufficiently small. In the
standard Ginzburg--Landau theory, this is only true near the critical
temperature. It is, however, plausible to think of a zero-temperature effective
field theory for the superconducting phase. We therefore understand our
Lagrangian as such an effective expansion in terms of the Nambu--Goldstone
modes \cite{Casalbuoni:1999wu,Son:1999cm} generalized by inclusion of modes of
the modulus of the order parameter \cite{Rischke:2000qz,Aitchison:1994hu}. In
ordinary superconductivity, the Nambu--Goldstone mode is the
Bogolyubov--Anderson mode, and the modulus mode is the Abrahams--Tsuneto mode
\cite{Abrahams:1966ab}.

Our last remark points to the above calculated masses of gluons generated by
the Higgs mechanism. To specify the scale of the masses one would have to know
the normalization coefficients $\alpha_{e,m}$. These are unknown parameters of
the effective theory and have to be determined from the matching with the
microscopic theory. At zero temperature, they are roughly
\[\alpha_{e,m}\propto\mu^2/\phi^2,\]
and as a result, both electric and magnetic masses are found to be of order
$g\mu$, where $\mu$ is the baryon chemical potential. Their physical origin is,
however, very different. The electric (Debye) mass is non-zero even in the
normal state i.e., above the critical temperature, due to polarization effects
in the quark medium. On the other hand, the magnetic (Meissner) mass arises
purely as a consequence of the spontaneous symmetry breaking. It is thus zero
at the critical point and increases as the temperature is lowered, to become
roughly equal in order of magnitude to the Debye mass at $T=0$.

Unfortunately, this is not the end of the story. As pointed out by Rischke who
calculated the gluon masses microscopically for the two-flavor color
superconductor \cite{Rischke:2000qz}, the lowest order kinetic term alone does
not give correct ratios of gluon masses of different adjoint colors. It is
therefore not of much help to just try to adjust the normalization of the
kinetic term. As a remedy to this problem, it is necessary to make use of
higher order contributions to the gluon masses.

In the two-flavor color superconductor with a color-antitriplet condensate,
there is only one generically different higher order contribution that can
change the ratios of the gluon masses from those given by the lowest order
kinetic term (see Ref. \cite{Rischke:2000qz}, Eq. 153). This reflects the
symmetry of the problem: the order parameter (conventionally chosen to point in
the direction of the third color) leaves unbroken an $SU(2)$ subgroup of the
original color $SU(3)$. Under the unbroken subgroup, the gluons of colors 4--7
transform as a complex doublet and thus have to receive equal masses, possibly
different from the mass of gluon 8. The most general gluon mass matrix is thus
specified by two parameters.

In our case of a color-sextet condensate, the $SU(3)$ symmetry can be
completely broken and we thus expect that there are in general no relations
among the eight gluon masses. We do not go into details here, but just list the
kinetic terms of order four in the field $\Phi$, which give gluon mass ratios
different from the lowest order values:
\begin{gather*}
\left|\tr(\Phi^{\dagger}D_i\Phi)\right|^2,\\
\tr\left[(D_i\Phi)^{\dagger}(D^i\Phi)\Phi^{\dagger}\Phi\right],\\
\tr\left[\Phi^{\dagger}(D_i\Phi)\Phi^{\dagger}(D^i\Phi)\right]+\text{h.c.},
\end{gather*}
and analogously the terms contributing to the electric gluon masses.

In our Lagrangian the $SU(3)\times U(1)$ symmetry is realized linearly and
these terms are found `by inspection'. It would be appropriate to repeat the
analysis using the non-linearly realized effective Lagrangian along the lines
of \cite{Casalbuoni:2000cn} analyzing the color-antitriplet case. The kinetic
terms should follow from symmetry considerations, albeit again with
theoretically undetermined coefficients.

Finally we note that as the Debye masses of all gluons are non-zero in the
normal state, one might expect that in the superconducting phase they remain
non-zero even for those gluons which correspond to unbroken symmetries, in
contrast to the conclusions of the effective theory discussed. However, as
shown by Rischke for the two-flavor color superconductor, the ``unbroken''
electric gluons have, somewhat surprisingly, zero Debye mass at $T=0$. This is
because the quark colors they interact with are bound in the condensate and
hence there are no low energy levels to be excited by long-wavelength
chromoelectric fields.

This line of reasoning can be easily carried over to our case, since due to the
diagonal nature of the matrix $\Delta$, one can immediately check which quark
colors participate in the condensate. We thus conjecture that the naive
expectation that the Debye masses of the gluons of the unbroken symmetry are
zero, is correct at zero temperature, as long as the colors that the gluon
interacts with both have non-zero gap $\Delta_i$. This is the case, for
instance, for the gluons of the $SO(2)$ and $SO(3)$ ground state symmetries
discussed before (see paragraphs (b) and (c) above).

To provide a waterproof verification of this conjecture, on should carry out a
microscopic calculation similar to that of \cite{Rischke:2000qz}.

\section{Fermionic BCS-type description}
In the previous section we used an effective Higgs-like theory to treat the
kinematics of color superconductivity with a color-sextet condensate. The
construction of the effective Lagrangian is based solely on the $SU(3)\times
U(1)$ symmetry. Such an approach is thus pretty convenient to extract as much
information about the kinematics as possible, but fails to explain the very
fact of Cooper pair formation. To understand the dynamics of color
superconductivity, we need a microscopic description of the quark system.

As is well known from BCS theory of superconductivity, fermions (quarks in our
case) will tend to form Cooper pairs if there is an attractive effective
two-body interaction between them. As is usual in attempts to describe the
behavior of deconfined QCD matter, we employ the Nambu--Jona-Lasinio model and
look for the diquark condensate as a constant self-consistent solution to the
equations of motion.

Because the excitation spectrum of cold strongly coupled deconfined QCD matter
at moderate baryon density is not known, the effective quark--quark interaction
relevant for color superconductivity can only be guessed. In any case the
excitations of such a matter are of two sorts:
\begin{enumerate}
\item
Colored quasiparticles excited by the primary quantum fields with modified
dispersion laws.
\item
Collective excitations, which can be in principle both colored and colorless,
and are excited by the appropriate polynomials of the primary quantum fields.
\end{enumerate}
We want to argue in favor of possible existence of massive color-octet
spin-zero collective modes excited by two gluon operators
\cite{Hansson:1982dv}, the exchange of which produces the desired effective
four-quark interaction attractive in the color-sextet quark--quark channel. The
(naive) point is that the QCD-induced force between two gluons, which can in
general be in any of
$$8\otimes8=1\oplus8\oplus8\oplus10\oplus\overline{10}\oplus27,$$
is attractive is the color-octet spin-zero configuration.

Inspired by this argument, we choose for our NJL-type analysis a four-quark
interaction which mimics the exchange of an intermediate color-octet scalar
particle. As we note below, however, we could have as well included
interactions with Lorentz vectors or tensors. Nonetheless, the Lorentz
structure of the interaction does not play almost any role in our calculation,
and we therefore restrict to the single interaction term
\eqref{ScalarExchangeInt} suggested above.

Our effective Lagrangian for one massive quark flavor thus reads
\begin{equation}
\label{NJLlagrangian}
{\cal L}=\bar\psi(i\slashed\partial-m+\mu\gamma_0)\psi+
G(\bar\psi\vec\lambda\psi)^2,
\end{equation}
where the arrow over Gell-Mann $\lambda$-matrices implies appropriate summation
over adjoint $SU(3)$ indices. Otherwise, Lorentz and color indices are
suppressed.

We treat the model Lagrangian \eqref{NJLlagrangian} in the mean-field
approximation. As this is a standard way of dealing with NJL-type models, we
sketch only the main steps. Detailed account of the techniques used can be
found, for example, in the recent paper by Alford et al. \cite{Alford:2002rz}.

To extract the color-sextet condensate, we split our Lagrangian into a free and
interacting part ${\cal L}'_0$ and ${\cal L}'_{\text{int}}$, respectively,
\begin{multline*}
{\cal
L}'_0=\bar\psi(i\slashed\partial-m+\mu\gamma_0)\psi+{}\\
+\frac12\bar\psi\Delta(C\gamma_5)\bar\psi^T-
\frac12\psi^T\Delta^{\dagger}(C\gamma_5)\psi,
\end{multline*}
\begin{equation*}
{\cal L}'_{\text{int}}=-\frac12\bar\psi\Delta(C\gamma_5)\bar\psi^T+
\frac12\psi^T\Delta^{\dagger}(C\gamma_5)\psi+G(\bar\psi\vec\lambda\psi)^2,
\end{equation*}
where $\Delta$ is the desired gap parameter which, as shown in the preceding
section, can be sought in the form of a real diagonal non-negative matrix in
the color space. We introduce the standard Nambu--Gorkov doublet notation,
\[
\Psi(p)=\left(
\begin{array}{c}
\psi(p) \\ \bar\psi^T(-p)
\end{array}
\right),
\]
in which the calculation of the free propagator amounts to inverting a
$2\times2$ matrix,
\[
S^{-1}(p)=\left(
\begin{array}{cc}
\slashed p-m+\mu\gamma_0 & \Delta(C\gamma_5)\\
-\Delta^{\dagger}(C\gamma_5) & (\slashed p+m-\mu\gamma_0)^T
\end{array}
\right).
\]
The explicit form of the propagator has been given by several authors, see, for
instance, \cite{Pisarski:1999av,Huang:2001yw}.

In the mean-field approximation, $\Delta$ is determined from a single one-loop
Feynman graph. Regulating the quadratic divergence with a three-momentum cutoff
$\Lambda$ and evaluating explicitly the Wick-rotated integral over the temporal
component of the loop momentum, we finally arrive at the gap equation
\begin{equation}
\label{GapEquation} 1=\frac23G\int^{\Lambda}\frac{d^3\vec
p}{(2\pi)^3}\left(\frac1{E_+}+\frac1{E_-}\right),
\end{equation}
where $E_{\pm}$ represent the positive energies given by the dispersion
relations of the quasiquark excitations,
\[
E^2_{\pm}=\left(\sqrt{\vec p^2+m^2}\pm\mu\right)^2+|\Delta|^2.
\]

A few remarks to the gap equation \eqref{GapEquation} are in order. First, in
the loop integral, we have ignored a term proportional to $\mu\gamma_0$ which
generates the operator $\bar\psi(C\gamma_5)\gamma_0\bar\psi^T$ that breaks
Lorentz invariance. In fact, we should have expected such a term to appear,
since Lorentz invariance is explicitly broken by the presence of the chemical
potential in the Lagrangian \eqref{NJLlagrangian}. For our treatment of color
superconductivity at non-zero chemical potential to be fully consistent, we
would have to include such operators into our Lagrangian from the very
beginning and solve a coupled set of gap equations for both Lorentz invariant
and non-invariant condensates \cite{Buballa:2001wh}. Here, for the sake of
simplicity, we ignore this difficulty and neglect the secondary effects of
Lorentz-invariance breaking induced by the chemical potential.

Second, the gap equation \eqref{GapEquation} is understood as a matrix equation
in the color space. Its matrix structure is, however, trivial. In fact, we get
three separated identical equations for the diagonal elements
$\Delta_1,\Delta_2,\Delta_3$ of the gap matrix. This means that, at least at
the level of the mean-field approximation, our model favors an enhanced $SO(3)$
symmetry of the ground state --- the gaps for all three colors are the same.
This is apparently not a peculiar consequence of our particular choice of
interaction in \eqref{NJLlagrangian}, but holds for any $SU(3)$-invariant
four-fermion interaction. The only effect of adding also the Lorentz vector or
tensor channel interactions, for example, would be in the modification of the
effective coupling constant $G$. The Lorentz structure of the interaction does
not play any role and the resulting form of the gap equation is a consequence
of the identity $\vec\lambda\Delta\vec\lambda^T=4\Delta/3$, which holds for any
diagonal matrix $\Delta$. We will return to the discussion of this point in the
next section where we will comment on a correspondence between the bosonic and
fermionic approaches.

Third, the extension of the gap equation to non-zero temperatures is easy. We
can either first calculate the thermodynamical potential $\Omega$ and then
minimize it with respect to $\Delta$ or, alternatively, proceed in the same
manner as before and derive a self-consistency condition for the thermal Green
function \footnote{Once we have found the gap equation, we can obtain the
thermodynamical potential by integrating it over the gap parameter (for further
details see \cite{Zhou:2002fy}).}. Performing the sum over Matsubara
frequencies in the last step, the result is
\[
1=\frac23G\int^{\Lambda}\frac{d^3\vec
p}{(2\pi)^3}\left(\frac1{E_+}\tanh\frac{E_+}{2T}
+\frac1{E_-}\tanh\frac{E_-}{2T}\right).
\]
This gap equation can be used for the study of temperature dependence of the
gap and, in particular, for finding the critical temperature at which the $SU(3)$
symmetry is restored \cite{Pisarski:1999av}.

\section{Summary and discussion}
Let us briefly summarize our results. First we developed the Higgs mechanism
for a color sextet and found out that although the underlying symmetry allows
for a complete spontaneous breakdown, for a generic quartic scalar potential
the pattern $SU(3)\to SO(3)$ is preferred.

After then, we used the Nambu--Jona-Lasinio model to calculate the gaps
$\Delta_1,\Delta_2,\Delta_3$ self-consistently in the mean-field approximation
and our result was in accord with the preceding Higgs-type analysis.

This is, of course, not only a coincidence, but follows from a general
correspondence between four-fermion-interaction models and linear sigma models
provided by the Hubbard--Stratonovich transformation.

Let us sketch the main idea. In the path integral formalism, one first
introduces an auxiliary scalar integration variable which has no kinetic term
and couples to the fermion via the Yukawa interaction. The action now becomes
bilinear in the fermion variables and one can integrate them out explicitly. The
logarithm of the fermion determinant gives rise to a kinetic term of the scalar
field and the model hence becomes equivalent to the linear sigma model, up to a
choice of the renormalization prescription \cite{Eguchi:1976iz}.

In terms of the NJL model the interpretation of the correspondence is a little
bit different. Here one cannot carry out the usual renormalization program and
the choice of an ultraviolet regulator becomes physically significant. So in
the effective scalar field action the operators with dimension four or less are
dominant since they are generated with divergent coefficients. The quadratic
divergences cancel due to the gap equation in the underlying NJL model but the
logarithmic ones remain \cite{Ripka:1997zb}.

One thus receives an a posteriori justification for the choice of the linear
sigma model as the starting point for the Higgs-type analysis in subsection
II.C. On the other hand, one should bear in mind that these conclusions are
valid only in the mean-field approximation that we employed.

In terms of the effective scalar field $\Phi$, the true vacuum is determined by
the absolute minimum of the full quantum effective potential which is no longer
restricted to contain operators of dimension four or less.

In the NJL model, going beyond the mean-field approximation \footnote{Formally,
we imagine this as adding two- and more-loop graphs, with full fermion
propagators inserted in place of the free ones, to the right hand side of the
gap equation \eqref{GapEquation}. Physically this means that the effective
coupling constant $G$ may depend on the order parameter $\Delta$
\cite{Tsuneto:1998ts}.} could destroy the simple structure of the one-loop gap
equation \eqref{GapEquation}. Generally, the resulting set of algebraic
equations for $\Delta_1,\Delta_2,\Delta_3$ must be permutation invariant since
permutations of diagonal elements of the matrix $\Delta$ belong to the symmetry
group $SU(3)$ of the theory. For four-fermion interactions the $SU(3)$
structure of an arbitrary Feynman graph can be investigated making use of the
Fierz identities in the color space. One gets three coupled, but still rather
simple equations for the three gaps. It is then perhaps a matter of numerical
calculations to decide whether these equations possess asymmetric solutions and
whether they are more energetically favorable than those with
$\Delta_1=\Delta_2=\Delta_3$.

We suspect that asymmetric solutions implying a complete breakdown of the
$SU(3)\times U(1)$ symmetry can also be obtained from interactions that mimic
many-body forces (six-fermion or more). The correspondence with linear sigma
model via the Hubbard--Stratonovich transformation is then lost and it could
hopefully suffice to stay at the level of the mean-field approximation, thus
requiring much less manual work than in the previous case.

Investigations in the two directions mentioned above are already in progress.

\begin{acknowledgments}
The authors are greatly indebted to Dirk Rischke for his kind and insightful
remarks which resulted in the discussion in subsection II.E. They are also
grateful to Micaela Oertel and Michael Buballa for critical comments on an
early draft of the paper. T.B. would like to thank W. Grimus for bringing the
reference \cite{Schur:1945ab} to his attention. J.H. acknowledges with pleasure
Uwe-Jens Wiese for several useful and pleasant discussions.

This work was supported in part by grant GACR 202/02/0847. The work of T.B. was
also in part supported by a graduate fellowship of the Faculty of Mathematics
and Physics, Charles University.
\end{acknowledgments}

\end{document}